\documentclass[a4paper,11pt]{article}

\usepackage{jinstpub} 

\usepackage{threeparttable}

\title{\boldmath Sub-pixel Response of Double-SOI Pixel Sensors\\for X-ray Astronomy}

\author[a,1]{Kouichi~Hagino,\note{Corresponding author.}}
\author[a]{Kousuke~Negishi,}
\author[a]{Kenji~Oono,}
\author[a]{Keigo~Yarita,}
\author[a]{Takayoshi~Kohmura,}
\author[b]{Takeshi~G.~Tsuru,}
\author[b]{Takaaki~Tanaka,}
\author[b]{Sodai~Harada,}
\author[b]{Kazuho~Kayama,}
\author[c]{Hideaki~Matsumura,}
\author[d]{Koji~Mori,}
\author[d]{Ayaki~Takeda,}
\author[d]{Yusuke~Nishioka,}
\author[d]{Masataka~Yukumoto,}
\author[d]{Kohei~Fukuda,}
\author[d]{Takahiro~Hida,}
\author[e]{Yasuo~Arai,}
\author[f]{Ikuo~Kurachi,}
\author[g]{and Shunji~Kishimoto}

\affiliation[a]{Department of Physics, School of Science and Technology, Tokyo University of Science, \\Noda, Chiba 278-8510, Japan}
\affiliation[b]{Department of Physics, Faculty of Science, Kyoto University, \\Sakyo, Kyoto 606-8502, Japan}
\affiliation[c]{Physics and Mathematics of the Universe (Kavli IPMU, WPI), The University of Tokyo, \\Kashiwa, Chiba 277-8583, Japan}
\affiliation[d]{Department of Applied Physics, Faculty of Engineering, University of Miyazaki, \\Miyazaki, Miyazaki 889-2155, Japan}
\affiliation[e]{Institute of Particle and Nuclear Studies (IPNS), High Energy Accelerator Research Organization (KEK), \\Tsukuba, Ibaraki 305-0801, Japan}
\affiliation[f]{Department of Advanced Accelerator Technologies (AAT), High Energy Accelerator Research Organization (KEK), \\Tsukuba, Ibaraki 305-0801, Japan}
\affiliation[g]{Institute of Materials Structure Science (IMSS), High Energy Accelerator Research Organization (KEK), Tsukuba, Ibaraki 305-0801, Japan}

\emailAdd{hagino@rs.tus.ac.jp}

\abstract{
We have been developing the X-ray silicon-on-insulator (SOI) pixel sensor called XRPIX for future astrophysical satellites. XRPIX is a monolithic active pixel sensor consisting of a high-resistivity Si sensor, thin SiO$_2$ insulator, and CMOS pixel circuits that utilize SOI technology. Since XRPIX is capable of event-driven readouts, it can achieve high timing resolution greater than $\sim10{\rm ~\mu s}$, which enables low background observation by adopting the anti-coincidence technique. One of the major issues in the development of XRPIX is the electrical interference between the sensor layer and circuit layer, which causes nonuniform detection efficiency at the pixel boundaries. In order to reduce the interference, we introduce a Double-SOI (D-SOI) structure, in which a thin Si layer (middle Si) is added to the insulator layer of the SOI structure. In this structure, the middle Si layer works as an electrical shield to decouple the sensor layer and circuit layer. We measured the detector response of the XRPIX with D-SOI structure at KEK. We irradiated the X-ray beam collimated with $4{\rm ~\mu m\phi}$ pinhole, and scanned the device with $6{\rm ~\mu m}$ pitch, which is 1/6 of the pixel size. In this paper, we present the improvement in the uniformity of the detection efficiency in D-SOI sensors, and discuss the detailed X-ray response and its physical origins.
}

\keywords{X-ray detectors, Space instrumentation, Imaging spectroscopy}

\proceeding{The 9$^{\text{th}}$ International Workshop on Semiconductor Pixel Detectors for Particles and Imaging\\
December 10--14, 2018\\
Academia Sinica, Taipei, Taiwan}

\begin{document}
\maketitle
\flushbottom

\section{Introduction}
\label{sec:intro}
Broadband X-ray imaging spectroscopy is an important probe for non-thermal emissions from high-energy celestial objects, such as supernova remnants, galaxy clusters, and black holes. Although nonthermal emissions are dominant in the hard X-ray band above $10{\rm ~keV}$, wide coverage from a few keV is required by the broadband and time-variable nature of nonthermal emissions. In order to realize such observations, we propose a future X-ray mission, FORCE~\cite{Mori2016,Nakazawa2018}. The FORCE mission will realize broadband X-ray imaging spectroscopy from 1~keV to 80~keV with angular resolution higher than 15~arcsecond by a combination of lightweight Si supermirrors and stacked Si/CdTe semiconductor detectors located at the focal plane of the supermirrors. The focal plane detector is surrounded by well-type active shields, which are used as anti-coincidence counters to reject the in-orbit non-X-ray background with the anti-coincidence technique. Owing to the high angular resolution and anti-coincidence technique, the sensitivity of FORCE will be one order of magnitude better that previous broadband X-ray missions such as the HXI onboard Hitomi~\cite{Nakazawa2018a,Hagino2018} and NuSTAR~\cite{Harrison2013}.

We are developing X-ray silicon-on-insulator (SOI) pixel sensors called XRPIX as a focal plane detector~\cite{Arai2011,Tsuru2018}. XRPIX is a monolithic pixel sensor that consists of a thick, high-resistivity Si sensor and CMOS pixel circuits, sandwiching a thin SiO$_2$ insulator. This configuration enables low readout noise of a few electrons (rms) and a depletion layer thickness of a few hundred micrometers, both of which are essential to achieve broadband coverage from $1{\rm ~keV}$ to $40{\rm ~keV}$. Moreover, implementing a trigger function in the pixel circuits gives XRPIX a timing resolution better than $\sim10{\rm ~\mu s}$, which is indispensable for the anti-coincidence technique.

One of the major issues in development of XRPIX is poor X-ray detection efficiency at the pixel boundary. This issue was first reported in XRPIX1b by Matsumura~et~al.~(2014)~\cite{Matsumura2014}. Based on beam scanning experiments at the sub-pixel scale and semiconductor device simulations, it was revealed that the poor efficiency at the pixel boundary was caused by the local minima of electrostatic potential in the sensor layer~\cite{Matsumura2015}. Since the potential local minima is due to electrical interference between the pixel circuits and sensor layer, we developed new devices by re-arranging the pixel circuit to avoid interference. Negishi~et~al.~(2018) evaluated this new device XRPIX3b~\cite{Takeda2015} at the sub-pixel scale utilizing a synchrotron radiation facility~\cite{Negishi2018}. Although the detection efficiency at the two-pixel boundary improved in XRPIX3b from XRPIX1b, it remained below 80\% at the four-pixel boundary compared with that at the pixel center. In order to reduce the electrical interference between the pixel circuits and sensor layer, we developed a Double SOI (D-SOI) pixel sensor, in which a thin Si layer was added to the insulator layer~\cite{Miyoshi2013,Ohmura2016,Miyoshi2017}. With this new D-SOI sensor, we performed a beam scanning experiment at a synchrotron radiation facility, and evaluated its sub-pixel X-ray response. In this paper, we report on the detection efficiency and spectral response of the D-SOI pixel sensor XRPIX6D at the sub-pixel scale. We also discuss the possible physical origins of charge loss in the spectral shape found in this device.

\section{X-ray Beam Scanning Experiment of Double-SOI Sensor}
\begin{figure}[tbp]
\begin{minipage}{0.55\hsize}%
\centering
\includegraphics[width=\hsize]{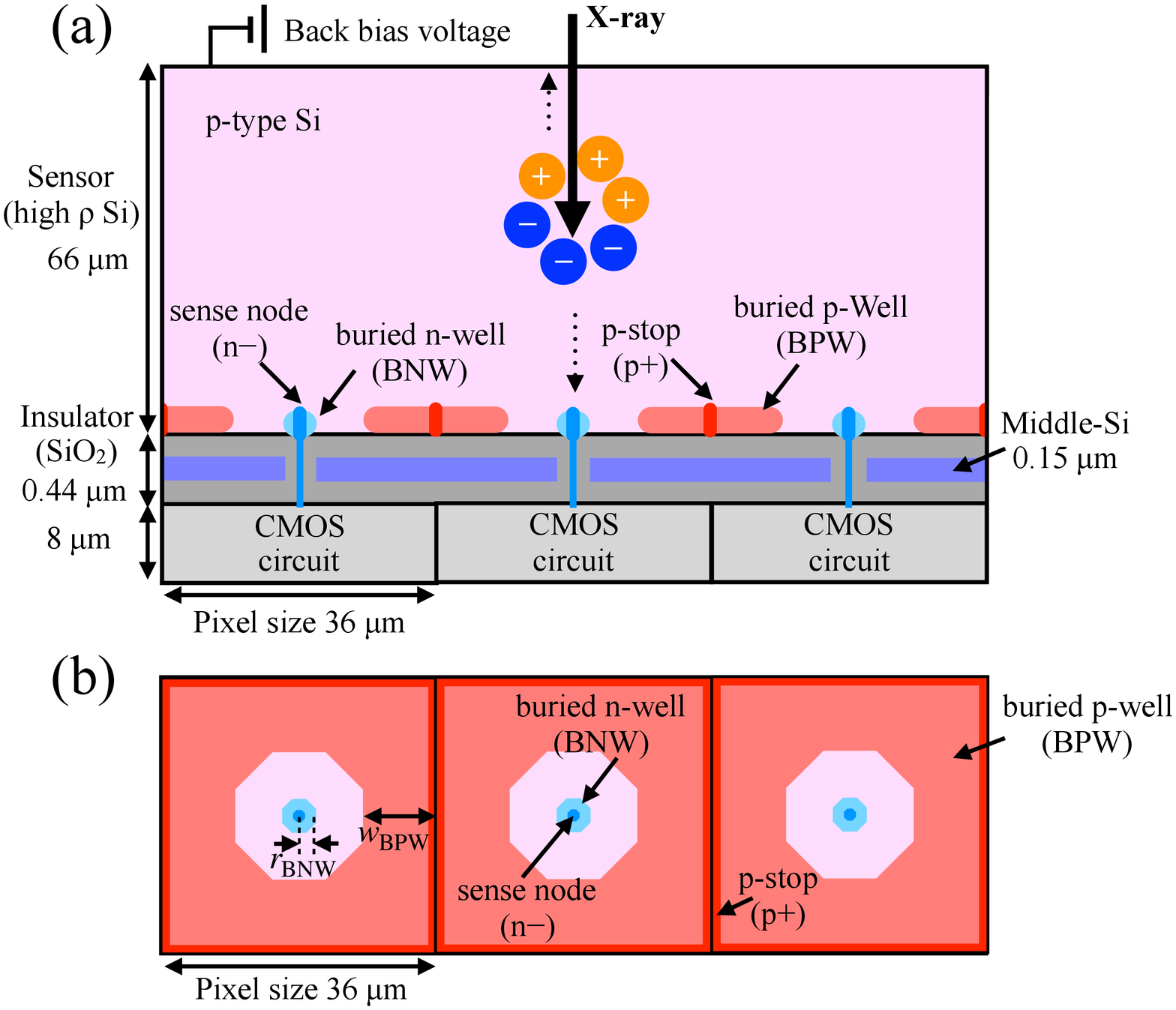}
\caption{(a) Vertical and (b) horizontal cross-sections of XRPIX6D.}
\label{fig:xrpix6D}
\end{minipage}%
\hspace{0.02\hsize}
\begin{minipage}{0.4\hsize}%
\makeatletter%
\def\@captype{table}%
\makeatother%
\centering
\caption{Specifications of XRPIX6D}
\smallskip
\begin{tabular}{|l|r|}
\hline
Parameters & Value\\
\hline
Sensor thickness & $66{\rm ~\mu m}$\\
Sensor resistivity & $1{\rm ~k\Omega~cm}$\\
Pixel size & $36{\rm ~\mu m}\times 36{\rm ~\mu m}$\\
BPW size $w_{\rm BPW}$ & $1.5,~3.5,~5.5,~7.5,$\\
					 & $9.5,~11.5,~13.5{\rm ~\mu m}$\\
\hline
\end{tabular}
\label{tab:spec}
\end{minipage}%
\end{figure}

We evaluated the sub-pixel X-ray response of the D-SOI pixel sensor called XRPIX6D~\cite{Takeda2019}. As shown in Figure~\ref{fig:xrpix6D}, an additional Si layer referred to as middle-Si was introduced into the insulator layer. When a negative bias voltage is applied, the middle-Si layer works as an electrostatic shield to decouple the sensor layer and pixel circuits. In addition, there are two more differences from the previous device XRPIX3b. The first one is the type of substrate. The sensor layer of XRPIX6D is p-type, while that of XRPIX3b is n-type. Thus, in XRPIX6D, electrons rather than holes are collected to the sense nodes. The second difference is the introduction of a floating p$^+$ implantation called p-stop and buried p-well (BPW) at the pixel boundary. This structure prevents the shorting out of pixels due to fixed charges in the SiO$_2$ layer. It will also be useful in terms of charge collection efficiency at the pixel boundary because it forms a potential barrier and pushes charge carriers to the sense nodes.

The specifications of XRPIX6D are listed in Table~\ref{tab:spec}. XRPIX6D has a relatively thin sensor layer with a thickness of $66{\rm ~\mu m}$ because of its relatively low resistivity of $1{\rm ~k\Omega}$. Such low resistivity requires as high as $\simeq400{\rm ~V}$ for a $200{\rm \textrm{-}\mu m}$ depletion layer. Full depletion is also necessary at energies of a few keV. Thus, a relatively thin sensor layer was adopted for this device. The pixel size is $36{\rm ~\mu m}$, which is slightly larger than those of XRPIX3b ($30{\rm ~\mu m}$). This device contains test element groups (TEGs) with different BPW size $w_{\rm BPW}$ ranging from $1.5{\rm ~\mu m}$ to $13.5{\rm ~\mu m}$. As shown in Figure~\ref{fig:xrpix6D}b, the BPW size $w_{\rm BPW}$ is defined as the width of the BPW in one pixel measured from the pixel boundary. Since the energy resolution (full-width half maximum of Mn K$\alpha$ line) was best at $w_{\rm BPW}=9.5{\rm ~\mu m}$ for the typical back bias voltage of $V_{\rm BB}=-100{\rm ~V}$, the experiment was performed for this TEG.

\begin{figure}[tbp]
\centering
\includegraphics[width=0.7\hsize]{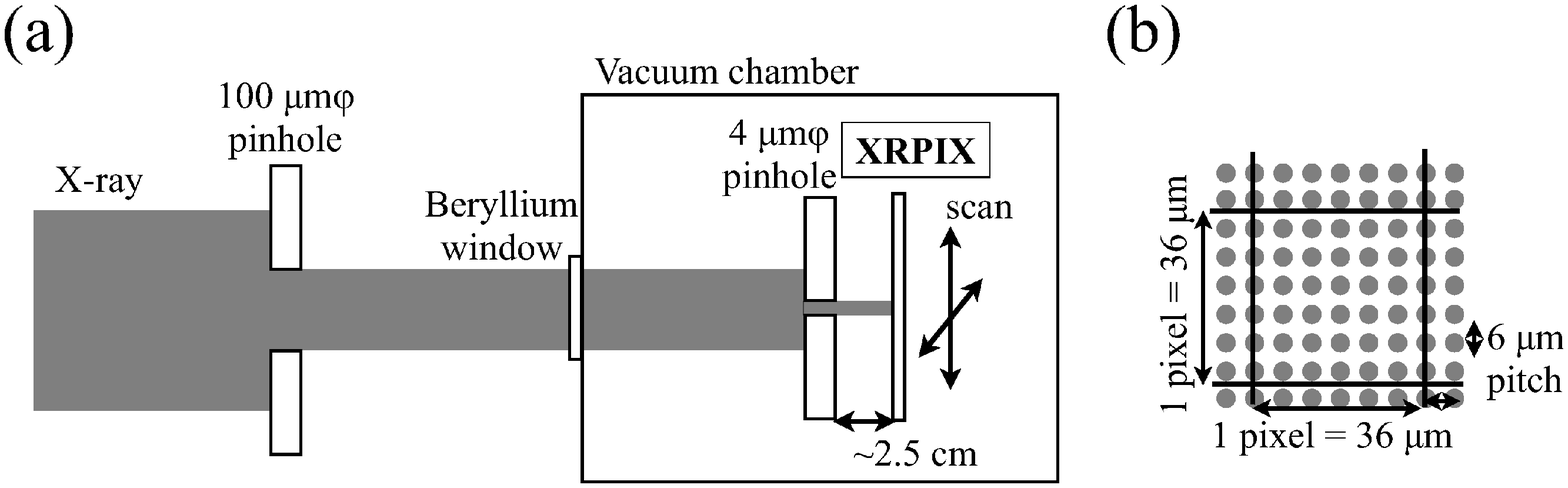}
\caption{(a) A schematic view of the experimental setup at the BL14A in the Photon Factory of KEK and (b) scanning procedure.}
\label{fig:setup}
\end{figure}

A beam scanning experiment of XRPIX6D was performed at the beamline BL14A of the synchrotron radiation facility, the Photon Factory of the High Energy Accelerator Research Organization (KEK) in Japan~\cite{Satow1989}. BL14A provides monochromatic X-ray beams in a wide energy range from $\sim6{\rm ~keV}$ to $\sim80{\rm ~keV}$. We used an X-ray beam of 6.0~keV in this experiment. The X-ray beam was collimated with $100{\rm ~\mu m}\phi$ and $4{\rm ~\mu m}\phi$ pinholes, and irradiated to the XRPIX6D located in a vacuum chamber, as shown in Figure~\ref{fig:setup}a. The $4{\rm ~\mu m}\phi$ pinhole was made of gold with a thickness of $90{\rm ~\mu m}$. The X-ray beam was irradiated on the back side of the device to avoid total ionization damage~\cite{Yarita2018}. XRPIX6D was set on the X-Z stage, and moved along vertical and horizontal directions with a step size of $6{\rm ~\mu m}$ as shown in Fig.~\ref{fig:setup}b.

The device was cooled down to $-70^\circ{\rm C}$ to reduce readout noise due to the leakage current. We applied a back bias voltage of $-100{\rm ~V}$, at which the sensor layer is fully depleted. In this experiment, we periodically read out the X-ray data from all pixels just like the readouts of charge-coupled devices, without using the trigger function. This is because we focused on the improvement from XRPIX3b, whose sub-pixel response was measured without using the trigger function.

The experimental data was analyzed using the method described in Ryu~et~al.~(2011)~\cite{Ryu2011} and Nakashima~et~al.~(2012)~\cite{Nakashima2012}. First, X-ray events were detected by searching pixels with pulse height above the predefined threshold called ``event threshold.'' Then, charge-sharing was judged by applying the ``split threshold'' to the adjacent pixels. According to the number of adjacent pixels exceeding the split threshold, the X-ray events were classified as ``single-pixel events,'' ``double-pixel events,'' etc.

\section{Sub-pixel Response of XRPIX6D}
\begin{figure}[tbp]
\centering
\includegraphics[width=0.95\hsize]{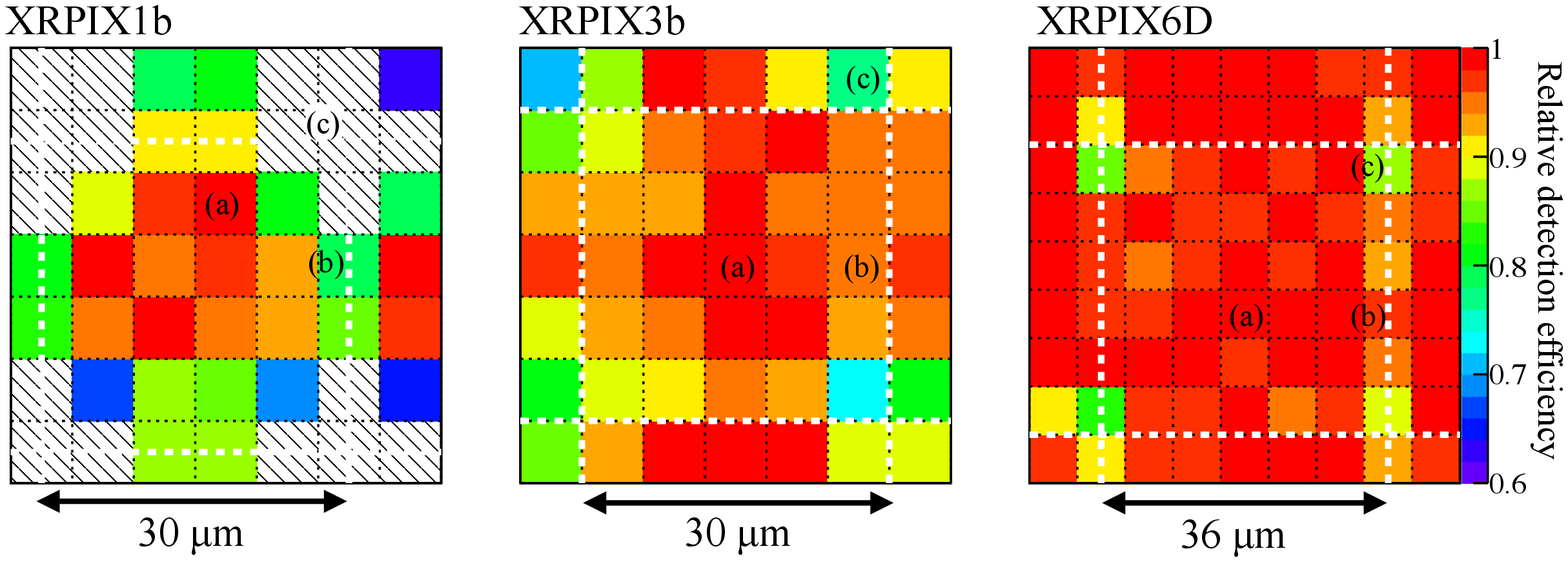}
\caption{Two-dimensional map of relative detection efficiency of XRPIX1b, XRPIX3b, and XRPIX6D at the sub-pixel scale. The shaded regions indicate positions with the relative detection efficiency below 0.6. As references, the estimated pixel boundaries are shown as white dotted lines.}
\label{fig:uniformity}
\end{figure}

Two-dimensional X-ray count map of XRPIX6D for $6.0{\rm ~keV}$ X-rays at the sub-pixel scale is shown in Figure~\ref{fig:uniformity}. The count rate in this map was normalized with the maximum count rate around the pixel center. This map is interpreted as a map of the relative detection efficiency at the sub-pixel scale. Those of XRPIX1b for 8.0~keV and XRPIX3b for 5.0~keV are also shown for comparison. The detection efficiency at the pixel boundary clearly improved in XRPIX6D compared with the previous devices. For a more quantitative comparison, ratios of the efficiencies at the pixel boundaries labeled as (b) and (c) in the map to the pixel center labeled as (a) are listed in Table~\ref{tab:uniformity}. Since it is difficult to determine the irradiation position at the sub-pixel scale, positions (a), (b), and (c) are not exactly the same among the three devices. The uncertainty of the estimated pixel boundary in the experimental data is similar to the scan pitch of $\sim6{\rm ~\mu m}$. Although there was no significant improvement at the two-pixel boundary from XRPIX3b, the efficiency at the four-pixel boundary significantly improved in XRPIX6D from XRPIX3b.

\begin{table}[tbp]
\centering
\caption{Detection efficiencies at two-pixel and four-pixel boundaries}
\begin{threeparttable}
\begin{tabular}{|c|c|c|c|}
\hline
Device & Two-pixel boundary\tnote{1} & Four-pixel boundary\tnote{1} & \# of events@pixel center\\
\hline
XRPIX1b (8.0~keV) & $81.1 \pm 2.8\%$ & $22.4 \pm 1.2\%$ & $\simeq2000$\\
XRPIX3b (5.0~keV) & $95.7 \pm 2.2\%$ & $76.3 \pm 1.9\%$ & $\simeq3000$\\
XRPIX6D (6.0~keV) & $96.1 \pm 2.4\%$ & $86.8 \pm 2.1\%$ & $\simeq3000$\\
\hline
\end{tabular}
\begin{tablenotes}\footnotesize
\item[1] The errors indicate $1\sigma$ statistical uncertainties.
\end{tablenotes}
\end{threeparttable}
\label{tab:uniformity}
\end{table}
\begin{figure}[tbp]
\centering
\includegraphics[width=0.5\hsize]{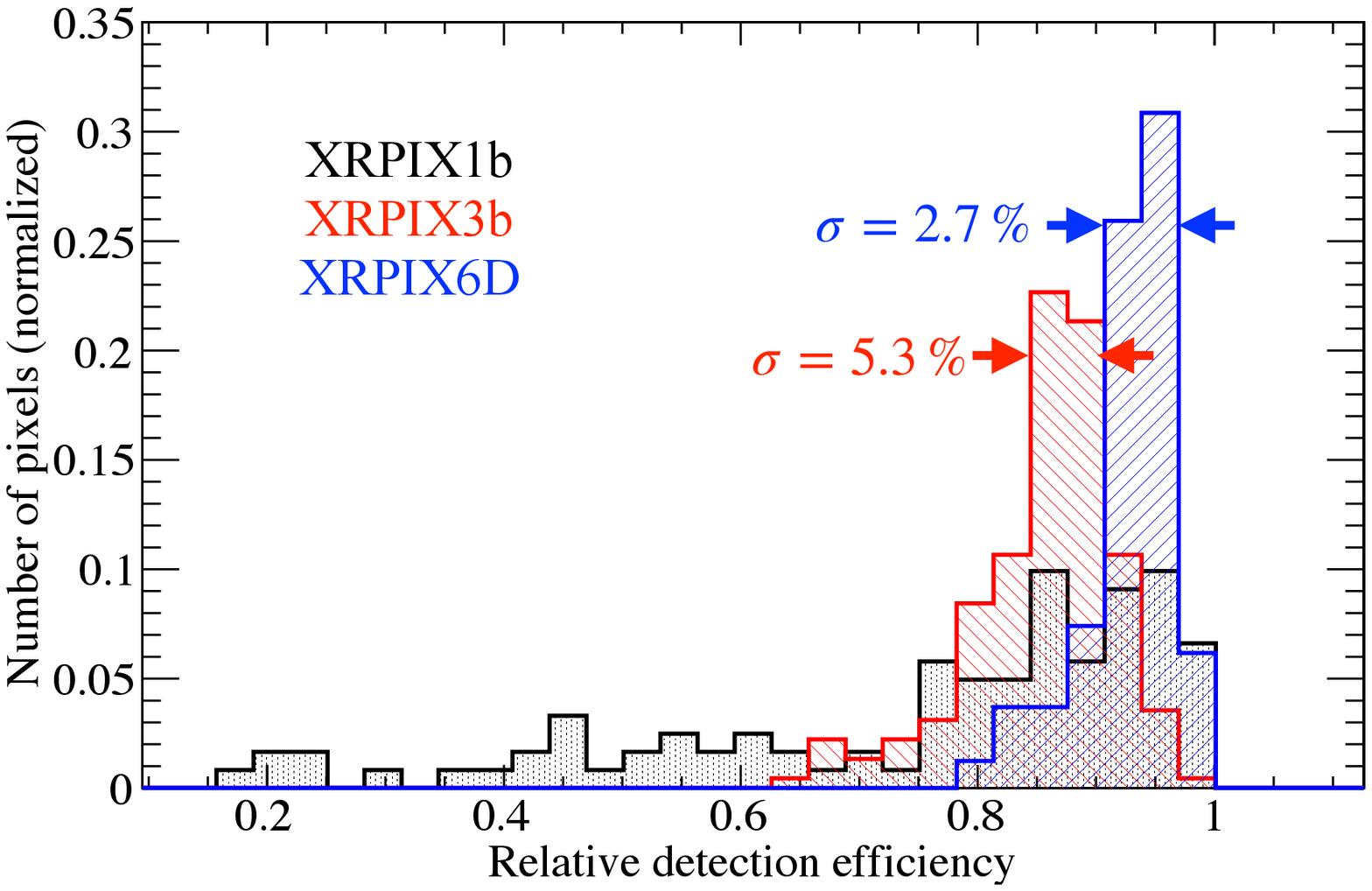}
\caption{One-dimensional histograms of the detection efficiency of each sub-pixel position. Y-axis is normalized so that the area of each histogram equals to unity.}
\label{fig:1dhist}
\end{figure}

The uniformity of all sub-pixel positions in the two-dimensional map (Figure~\ref{fig:uniformity}) was evaluated by plotting one-dimensional histograms of the efficiency at each position, as shown in Figure~\ref{fig:1dhist}. A tail-like structure below $\sim0.8$ in XRPIX1b corresponds to the low-efficiency regions at the pixel boundary in the two-dimensional map. For XRPIX3b and XRPIX6D, there are almost no tail-like structures except those corresponding to the four-pixel boundary, demonstrating the improved uniformity of detection efficiency. The standard deviations of the histograms of XRPIX3b and XRPIX6D are 5.3\% and 2.7\%, respectively. Since the standard deviation of 2.7\% in XRPIX6D is consistent with the statistical uncertainty of the number of counts, the detection efficiency at almost all sub-pixel positions except the four-pixel boundaries is uniform within the statistical uncertainty.

\begin{figure}[tbp]
\centering
\includegraphics[width=\hsize]{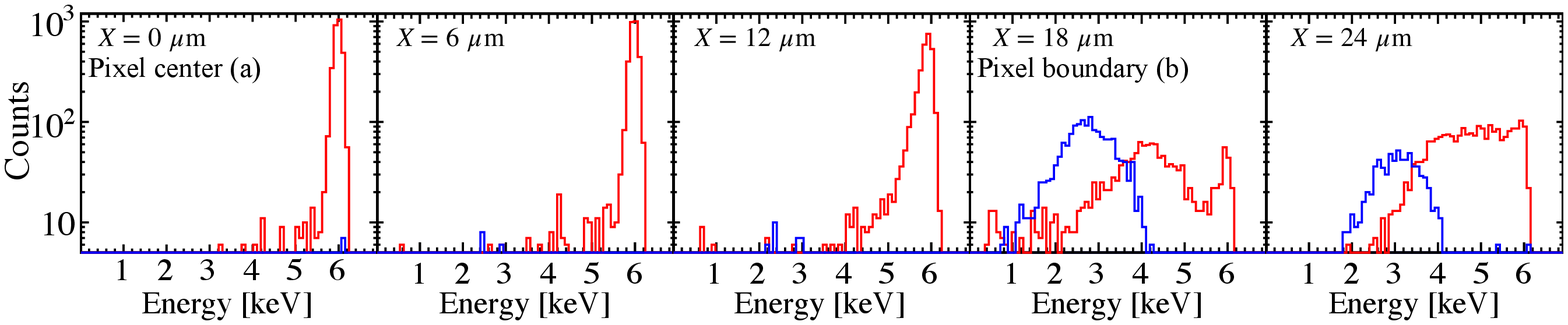}
\caption{Energy spectra of 6.0~keV X-ray obtained with XRPIX6D in the beam scanning experiment. Spectra of single-pixel events are shown in red lines, while those of double-pixel events are in blue lines. Here, $X$ is the horizontal axis at the sub-pixel scale measured from the pixel center (a).}
\label{fig:spec}
\end{figure}

The sub-pixel uniformity of the energy spectra was also evaluated. Figure~\ref{fig:spec} shows the spectra of each sub-pixel position. At the pixel center labeled as (a), almost all the charges produced by the 6.0~keV X-ray was collected by a single pixel. On the other hand, at the pixel boundary labeled as (b), a large amount of charge was split into one of the adjacent pixels, resulting in double-pixel events. In the spectra of the double-pixel events, all the split charges exceeding the split threshold were merged to correct the charge-sharing with adjacent pixels.

In the figure, two features can be seen. One is the tail structure in the spectra of single-pixel events, which increases as the irradiation position approaches the pixel boundary. Since the split threshold was set to be $0.18{\rm ~keV}$, the tail structure at $X=12{\rm ~\mu m}$ cannot be attributed to just the charge-sharing below the split threshold. The other feature is the lower pulse height of double-pixel events, which is less than half that of single-pixel events. Both of these results indicate that part of charge is lost at positions close to the pixel boundary.

Compared with the spectra with irradiation of Mn~K$\alpha$ at $5.90{\rm ~keV}$ from $^{55}$Fe on the full imaging area of the TEG with $w_{\rm BPW}=9.5{\rm ~\mu m}$, the spectral performance at the pixel center (a) is much better. The energy resolution at the pixel center was $220{\rm ~eV}$ in full-width half maximum (FWHM) at 6.0~keV, while that in the full imaging area was 290~eV (FWHM) at 5.90~keV. This demonstrates that the energy resolution decreased in the spectra of the full imaging area owing to the tail structure near the pixel boundary. Thus, investigating and resolving the charge loss in the spectral shape is essential to improve spectral performance.

\section{Cause of Charge Loss in the Spectral Shape}
\subsection{Investigations with Experimental Results}
\begin{figure}[tbp]
\centering
\includegraphics[width=0.35\hsize]{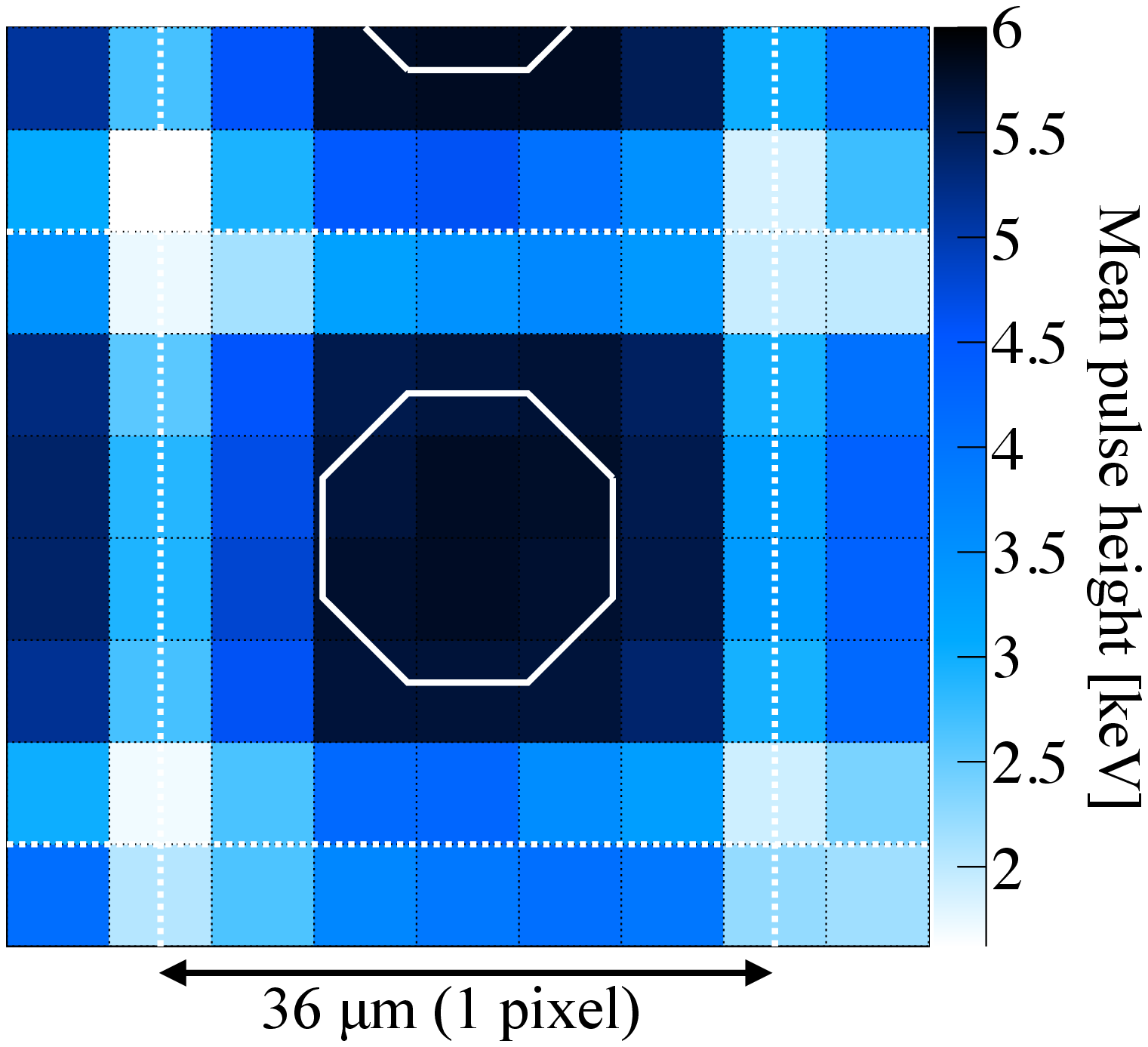}
\caption{Two-dimensional map of mean pulse height. The octagonal white solid lines indicate the inner edge of BPW.}
\label{fig:meanph}
\end{figure}

\begin{figure}[tbp]
\centering
\begin{minipage}[t]{0.48\hsize}
\centering
\includegraphics[width=\hsize]{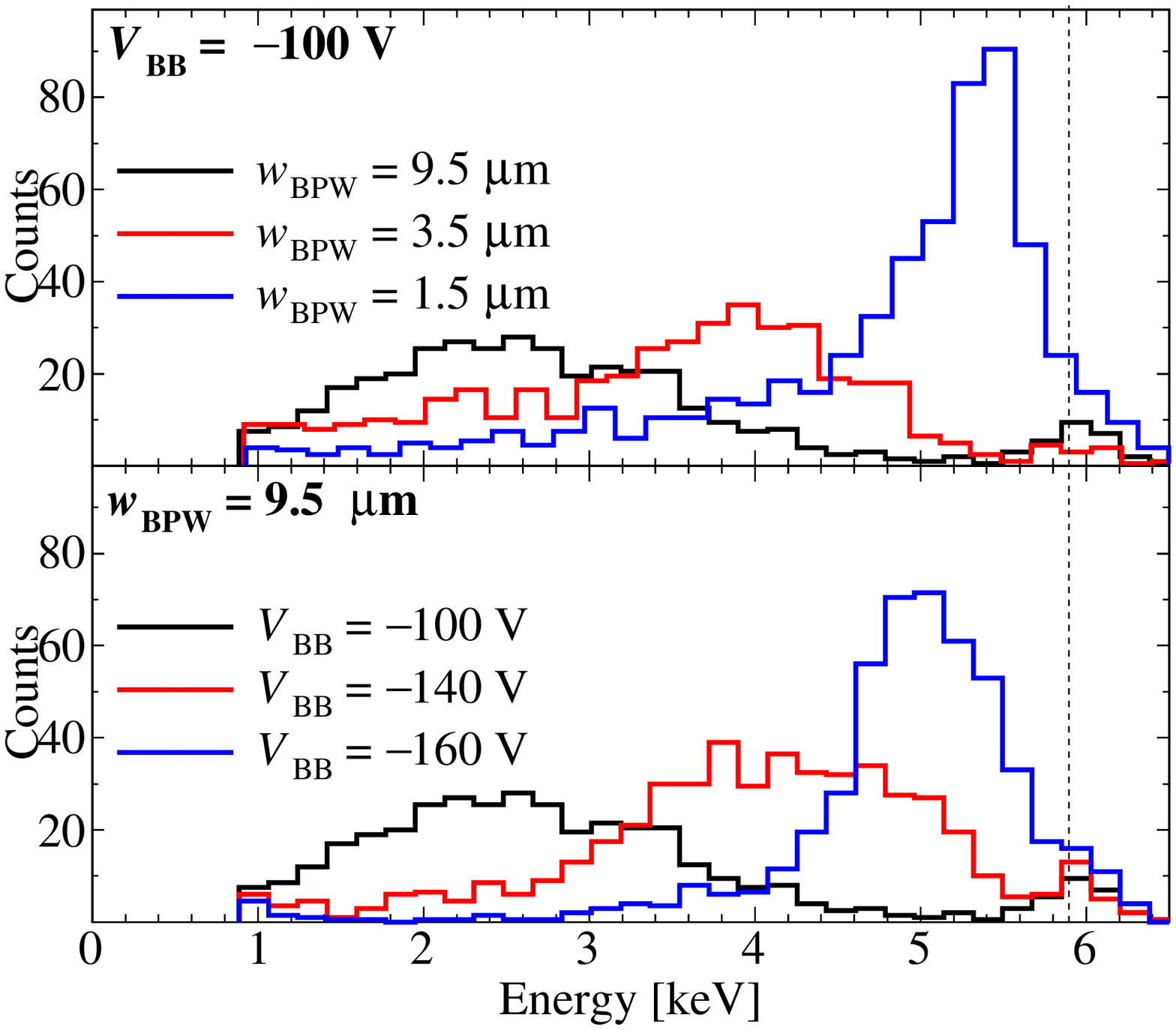}
\caption{Energy spectra of double-pixel events with an irradiation of Mn K$\alpha$ ($5.90{\rm ~keV}$) from $^{55}$Fe in the full imaging area. Top and bottom panels show the dependence on $w_{\rm BPW}$ and $V_{\rm BB}$, respectively.}
\label{fig:specs_bpw}
\end{minipage}
\hspace{0.02\hsize}
\begin{minipage}[t]{0.48\hsize}
\centering
\includegraphics[width=\hsize]{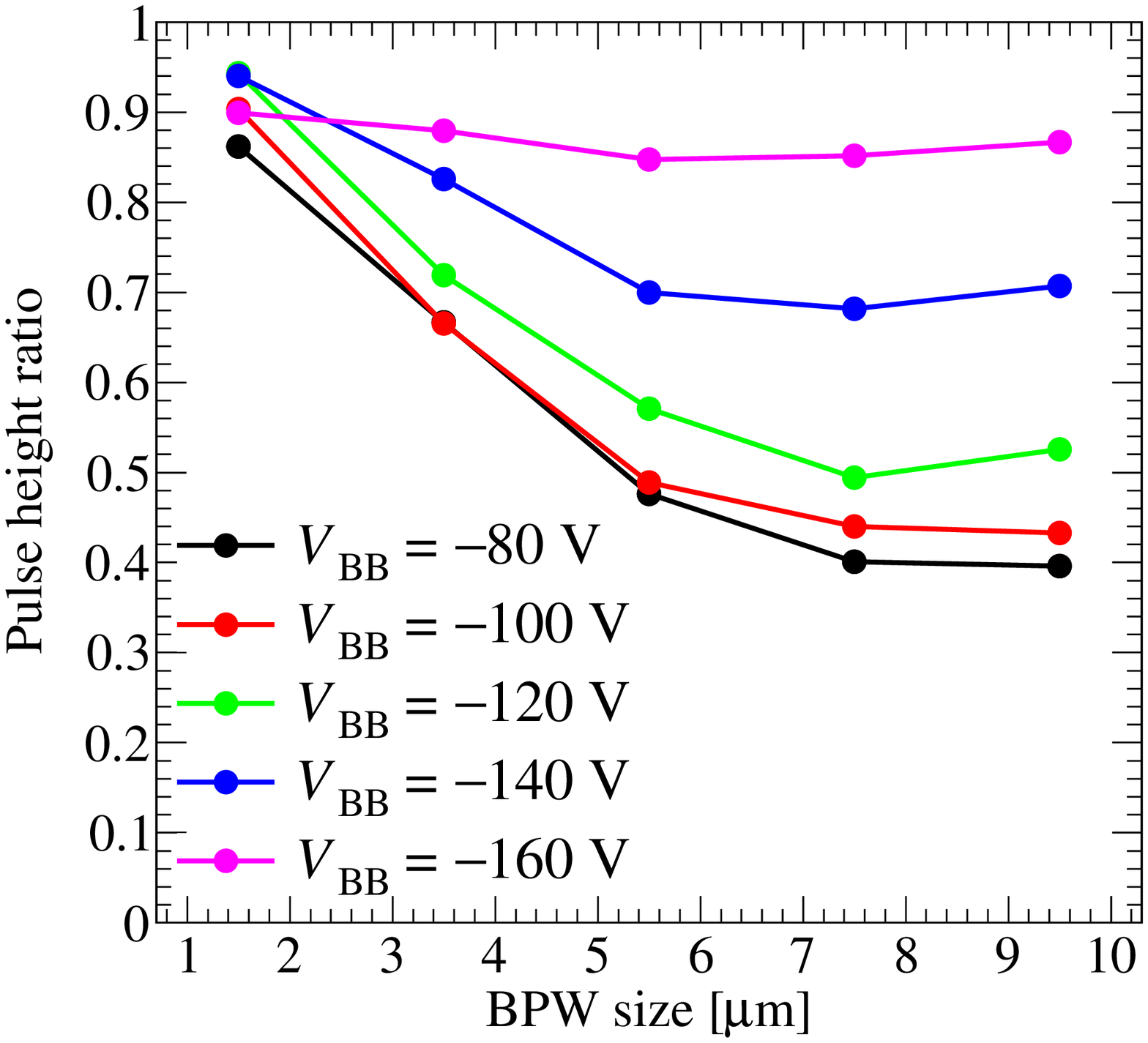}
\caption{Dependence of the pulse height ratio on BPW size $w_{\rm BPW}$ for different back bias voltages $V_{\rm BB}$.}
\label{fig:cce_bpw}
\end{minipage}
\end{figure}

To investigate the cause of the charge loss in the spectra, we quantitatively evaluated the amount of charge loss by calculating the mean pulse height of the spectra. Figure~\ref{fig:meanph} shows the dependence of the mean pulse height on the sub-pixel position. There is almost no charge loss at the positions surrounded by the inner edge of the BPW. Also, the mean pulse height decreased as the position became closer to the pixel boundary. Therefore, the charge loss seems to occur under the BPW.

Since the charge loss is supposed to be related to the BPW, the dependence of the charge loss on BPW size was measured by performing an additional experiment. We irradiated an X-ray of Mn K$\alpha$ from $^{55}$Fe on all TEGs with different BPW sizes. Figure~\ref{fig:specs_bpw} shows examples of the obtained spectra of double-pixel events. The charge loss clearly decreased with smaller BPW size and higher back bias voltage. Thus, we plotted the pulse height ratios of the double-pixel events to single-pixel events as a function of BPW size for different back bias voltages (figure~\ref{fig:cce_bpw}). This value would be a good indicator of the charge collection efficiency at the pixel boundary. The figure shows the strong dependence of the pulse height ratio on BPW size $w_{\rm BPW}$ and back bias voltage $V_{\rm BB}$. A larger amount of charge is lost under the condition of larger $w_{\rm BPW}$ and lower $V_{\rm BB}$. Therefore, the BPW must play a significant role in the charge loss. Note that the best energy resolution of single-pixel events was achieved at $w_{\rm BPW}=9.5{\rm ~\mu m}$ despite the large charge loss at this BPW size. This is because the charge loss in double-pixel events does not affect the core of the line profile, but only the amount of the tail structure.

\subsection{Investigations with Device Simulations}
Since it is difficult to investigate electric field structure and carrier transport in the sensor with only experimental data, we performed two-dimensional simulations of electrostatic potential in XRPIX6D using the semiconductor device simulator HyDeLEOS, which is a part of the TCAD system HyENEXSS~\cite{TCAD}. We implemented all the structures such as the sense nodes, p-stops, BPWs, BNWs, and middle-Si layers. We also set the positive fixed charge of $2\times10^{11}{\rm ~cm^{-2}}$ in the SiO$_2$ insulator layer according to previous work~\cite{Matsumura2015}.

\begin{figure}[tbp]
\centering
\includegraphics[width=0.75\hsize]{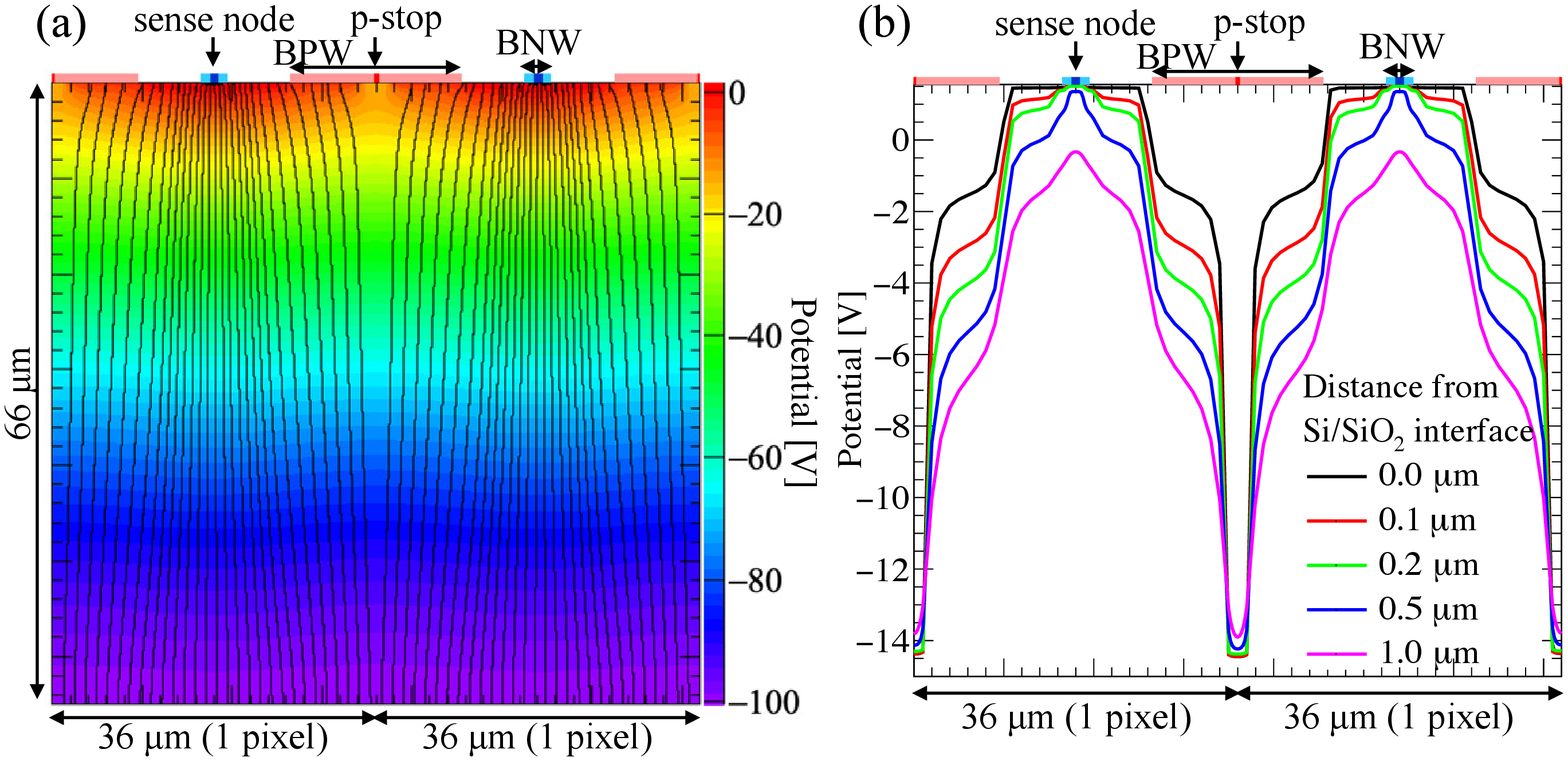}
\caption{Electrostatic potential (color) and electric field lines (black lines) in the (a) sensor layer, and (b) one-dimensional distributions of the potential near the Si/SiO$_2$ interface.}
\label{fig:potential}
\end{figure}

The potential map and electric field lines calculated by the simulations with back bias voltage of $V_{\rm BB}=-100{\rm ~V}$ and BPW size of $w_{\rm BPW}=9.5{\rm ~\mu m}$ are shown in Figure~\ref{fig:potential}a. Although the issue of poor detection efficiency in XRPIX1b was due to the local minimum of the electrostatic potential in the sensor layer, the simulated potential map in XRPIX6D showed no local minimum. Figure~\ref{fig:potential}b shows slices of the potential map near the Si/SiO$_2$ interface, where the local minimum existed in XRPIX1b. The potential at the Si/SiO$_2$ interface monotonically increased from the pixel boundary to the sense nodes without any local minimum structure. Thus, the charge loss issue was not caused by the potential local minimum, but possibly by carrier trapping or recombination.


Considering the experimental evidence of the dependence on the BPW size and the simulated potential map, we hypothesized that the charge carrier produced by X-ray is trapped or recombined during the drift in regions located under the BPW near the Si/SiO$_2$ interface. According to the simulated electric field shown in Figure~\ref{fig:potential}a, electric field lines at the pixel boundary are connected to the Si/SiO$_2$ interface under the BPW. Thus, the charge carrier generated at the pixel boundary moves vertically toward the interface, and then horizontally drifts under the BPW along the interface to the sense node. Therefore, if the charge loss region exists under the BPW, then the charge carrier generated at the pixel boundary would be lost just like the experimental result.

In this hypothesis, the dependence on the BPW size and back bias voltage in Figure~\ref{fig:cce_bpw} can be qualitatively explained. With the high back bias voltage, drift time under the BPW would decrease because the drift velocity increased with stronger electric field, which would improve the charge collection efficiency. Similarly, with smaller BPW, the drift time would decrease, leading to improved charge collection. In this hypothesis, the simulated electric field provides insightful information on this charge loss issue. In the simulation, the strength of the horizontal electric field under the BPW is typically $E\sim10^3{\rm ~V~cm^{-1}}$, corresponding to the drift time of $t=w_{\rm BPW}/\mu E\sim{\rm ns}$. Even if the lower electron mobilities in SiO$_2$ ($\mu\sim 20{\rm ~cm^2~V^{-1}~s^{-1}}$)~\cite{Hughes1973} or Si/SiO$_2$ interface ($\mu\sim 10^{2\textrm{--}3}{\rm ~cm^2~V^{-1}~s^{-1}}$)~\cite{Fang1968} are considered, the drift time would still be less than $\sim100{\rm ~ns}$. Thus, the carrier lifetime should be shorter than $\sim100{\rm ~ns}$. This very short lifetime implies very high density of the trapping/recombination centers. It would be a key parameter for further investigation of the physical origins of the charge loss.

\section{Conclusions}
We have evaluated the X-ray response of the D-SOI X-ray pixel sensor ``XRPIX6D'' at the sub-pixel scale utilizing a synchrotron radiation facility. Owing to the electrostatic shielding with the middle-Si layer in the D-SOI structure, the detection efficiency at the four-pixel boundary improved from 76\% in XRPIX3b to 88\% in XRPIX6D. Despite this improvement, a new issue emerged with the spectra, in which $\sim50\%$ of charge was lost at the pixel boundary. An additional experiment and TCAD device simulations indicate that the charge loss is probably due to carrier trapping or recombination under the BPW near the Si/SiO$_2$ interface.


\acknowledgments
We acknowledge the valuable advice and great work by the personnel of LAPIS Semiconductor Co., Ltd. This study was supported by MEXT/JSPS KAKENHI Grant-in-Aid for Scientific Research on Innovative Areas 25109002 (Y.A.), 25109003 (S.K.), 25109004 (T.G.T., T.T., K.M., A.T., and T.K.), Grant-in-Aid for Scientific Research (B) 25287042 (T.K.), Grant-in-Aid for Young Scientists (B) 15K17648 (A.T.), Grant-in-Aid for Challenging Exploratory Research 26610047 (T.G.T.), and Grant-in-Aid for JSPS Fellows 18J01417 (H.M.). This study was also supported by the VLSI Design and Education Center (VDEC), the University of Tokyo in collaboration with Cadence Design Systems, Inc., and Mentor Graphics, Inc.


\bibliographystyle{JHEP}
\bibliography{report}








\end{document}